# A Roadmap for Electronic Grade 2-Dimensional Materials

Natalie Briggs[1,2,3,#] & Shruti Subramanian[1,2,#], Zhong Lin[2,4], Xufan Li[5]†, Xiaotian Zhang[1,2,3], Kehao Zhang[1,2], Kai Xiao[5], David Geohegan[5], Robert Wallace[6], Long-Qing Chen[1], Mauricio Terrones[1,2,4,7], Aida Ebrahimi[1,8,9], Saptarshi Das[10], Joan Redwing[1,2,3], Christopher Hinkle[11], Kasra Momeni[12], Adri van Duin[1,7,10,13,14], Vin Crespi[2,4], Swastik Kar[15], Joshua A. Robinson[1,2,3,*]

[1]Department of Materials Science and Engineering, The Pennsylvania State University, University Park, PA, 16802
[2]Center for 2-Dimensional and Layered Materials, The Pennsylvania State University, University Park, PA, 16802
[3]2-Dimensional Crystal Consortium, The Pennsylvania State University, University Park, PA, 16802
[4]Department of Physics, The Pennsylvania State University, University Park, PA, 16802
[5]Center for Nanophase Materials Sciences, Oak Ridge National Laboratory, Oak Ridge, TN, 37831
[6]Department of Materials Science & Engineering, University of Texas at Dallas, Richardson, TX, 75080
[7]Department of Chemistry, The Pennsylvania State University, University Park, PA, 16802
[8]Department of Electrical Engineering, The Pennsylvania State University, University Park, PA, 16802
[9]Department of Biomedical Engineering, The Pennsylvania State University, University Park, PA, 16802
[10]Department of Engineering Science & Mechanics, The Pennsylvania State University, University Park, PA, 16802
[11]Department of Electrical Engineering, University of Notre Dame, Notre Dame, IN, 46556
[12]Department of Mechanical Engineering, Louisiana Technical University, Ruston, LA, 71272
[13]Department of Mechanical and Nuclear Engineering, The Pennsylvania State University, University Park, PA, 16802
[14]Department of Chemical Engineering, The Pennsylvania State University, University Park, PA, 16802
[15]Department of Physics, Northeastern University, Boston, MA, 02115

\# - Equal contributors
† - Current address: Honda Research Institute USA, Inc., Columbus, OH, 43212
\* - Corresponding author: jrobinson@psu.edu

Since their modern debut in 2004, 2-dimensional (2D) materials continue to exhibit scientific and industrial promise, providing a broad materials platform for scientific investigation, and development of nano- and atomic-scale devices. A significant focus of the last decade's research in this field has been 2D semiconductors, whose electronic properties can be tuned through manipulation of dimensionality, substrate engineering, strain, and doping.[1–8] 2D semiconductors such as molybdenum disulfide ($MoS_2$) and tungsten diselenide ($WSe_2$) have dominated recent interest for potential integration in electronic technologies, due to their intrinsic and tunable properties, atomic-scale thicknesses, and relative ease of stacking to create new and custom structures. However, to go "beyond the bench", advances in large-scale, 2D layer synthesis and engineering must lead to "exfoliation-quality" 2D layers at the wafer scale. This roadmap aims to address this grand challenge by identifying key technology drivers where 2D layers can have an impact, and to discuss synthesis and layer engineering for the realization of electronic-grade, 2D materials. We focus on three fundamental areas of research that must be heavily pursued in *both experiment and computation* to achieve high-quality materials for electronic and optoelectronic applications. The document is organized as follows:

## 1. Grand Challenges | *Where will 2D materials make an impact?*

The quest for fundamental knowledge will always drive discoveries; however, this roadmap is not an exhaustive guideline for future research directions in 2D science. Rather, it serves to identify key technological directions and milestones therein, where 2D materials are expected to have a sizeable impact in the future. The roadmap also establishes metrics that help accomplish specific milestones for each such technology challenge. This section identifies key technology drivers based on United States Grand Challenges (as determined by various bodies of experts),[9–16] where 2D-based technologies are expected to play significant roles. Analysis of these technology drivers provides key metrics which must evolve towards meeting challenge-specific goals. For example, electronic materials will require strong emphasis on metrics such as scalability, sheet resistance, mobility, doping, and CMOS compatibility. For the case of a filtration membrane, in contrast, the focus will more likely be on aspects such as chemical stability, porosity, and toxicity. In other words, this section can be viewed as an attempt to bridge the gap between 2D materials and the technologies they will drive, by identifying various structural and functional metrics.

Large-scale manufacturing of 2D materials is at its infancy. Although a few vendors for 2D materials exist, the current largest customer base is composed of research and development laboratories, which is not enough to sustain the growth and advancement of 2D materials synthesis technologies. Commercial 2D materials can still be considered an early-to-mid-stage technology in the Gartner Hype cycle,[17] and are likely a decade from the initial technology trigger. To this end, industry remains in the early stages of research and development, with small start-up companies pushing the advancements and media hype fueling the broad spectrum of possibilities. The success of the 2D materials story will require a careful identification of key technology drivers that can sail up the initial hype, pass quickly through the peak of inflated expectations, consolidate efficiently with early adopters, and move effectively towards the plateau of productivity. The timescale over which the field matures will vary vastly based on the complexity of the technology. For instance, technologies that depend on 2D inks will mature quicker than those which require high-speed electronic switching. The following paragraphs address some of these technology drivers.

As is always the case for such reports, it is impossible to cover all topics, and an attempt has been made to focus on the most relevant and impactful drivers. Of course, the challenges of the field continuously evolve, and, as a result, this report reflects the impactful areas as determined at the point of writing. A range of resources are utilized to identify the possible technology drivers for 2D materials, as they pertain to the Grand Challenges, outlined here.[9–16]

### 1.1 High Performance and Energy Efficient Computing

In 2015, the National Strategic Computing Initiative (NSCI) was created to spearhead US leadership in the area of high-performance computing (HPC).[10] In this initiative, HPC refers to systems that, through a combination of *processing capability* and *storage capacity*, can solve computational problems that are far beyond the (then) current 10-petaflop systems. The initiative envisions the accelerated development of "exascale computing systems" that operate at a thousand petaflops ($10^{18}$ flops) to exabytes of data.[10] Over the next 15 years, it also intends to establish a roadmap for future HPC systems *beyond the limits of current semiconductor technology*, (*i.e.* more-than-Moore technology). It stresses, among other objectives, a holistic developmental approach that addresses several relevant factors including *networking technology*, *downward scaling*, and *workforce development*, and benefits all stakeholders, including the industrial and academic sectors.

In addition to the need for high-performance computing, recent decades have brought forth a need for larger quantities of data processing at increased rates while consuming less power, in smaller volumes. To date, this is achieved by an *evolutionary downsizing* of the basic active and passive building block microprocessors. However, progress towards exascale computing faces a number of material, architectural, fabrication, and integration challenges to meet the required reliability, speed, compatibility, and power consumption metrics.[16,18] In this context, emerging new approaches in materials and systems are critical to address the computing power, data-storage capacities, thermal management, on- and off-board communication speeds, reliability, and CMOS-processing compatibility. Of these, processing speed and thermal management are combined into the energy-delay benchmark. In 2015, Nikonov and Young benchmarked a number of such emerging technologies using tunneling, ferroelectric, magneto-electric and spin-torque technology, as per the 2011 ITRS roadmap for its 2018 node.[18] Figure 1(a) summarizes the switching-energy versus delay benchmark of a 32-Bit adder based on these technologies. Leading contenders among these are the van der Waals (vdW) field effect transistors (FETs), (also known as 2D material FETs[19]) which lie on the lowest energy-delay product curve operating on 10 fJ of energy at ~500ps delay. To further investigate the potential of 2D materials-based FETs, Sylvia *et al.*[20] performed quantum mechanical simulations for vdW FETs with a number of monolayer transition metal dichalcogenides (TMDs) and black phosphorous (BP) channels operating in the low-voltage regime for geometries corresponding to those of the 2019 node and the 2028 node of the 2013 ITRS.[20] The results indicate that as the gate length is reduced from 13.3 nm (2019 node) to 5.9 nm (2028 node), leakage current becomes a challenge, and TMDs with heavier effective masses benefit most from extreme scaling. While the ballistic current always reduces with increasing technology node, direct tunneling through the channel and backscattering from the channel affect the total current. Specifically, it was found that an optimum effective mass of 0.4 $m_0$ corresponding to that of $WSe_2$ that provides a maximum drive current for operation with $V_{DD}$ = 0.3 V. Thus, at the 2028 node, the single gate vdW FETs show competitive drive current and power density. Figure 1(b) summarizes the energy-delay metrics at the 2019 and 2028 nodes for various single and double-gated vdW FETs.[11] In conjunction with these predicted performances, it is therefore exceedingly important for researchers to develop **a range of scalable 2D materials with high-performance and low-power device technologies.** Incorporating these 2D materials can create systems which apply exaflops of computing power to exabytes of data over the next decade.

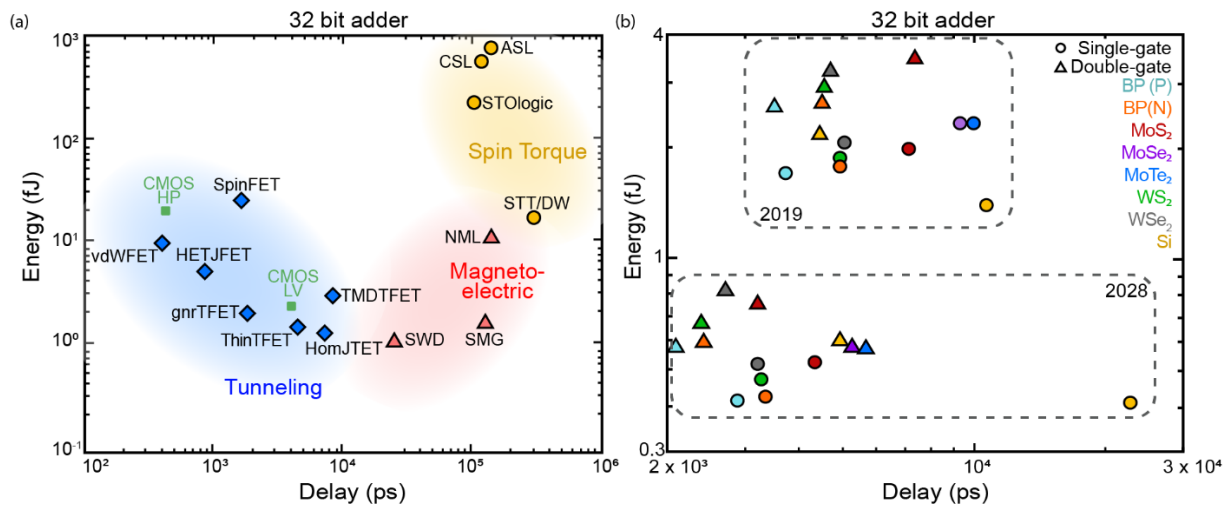

Figure 1: (a) Switching energy (in femtojoules) versus delay (in picoseconds) of a 32-bit adder for various emerging technologies (adapted from reference 20), which shows the promise of field effect transistors (FETs) based on 2D materials, i.e. vdW FETs, which lie at a low value of the energy-delay product metric. (b) Switching energy versus delay for 32-bit adders using various 2D materials for two technology nodes, namely the 2019 and the 2028 node (adapted from reference 18).

**1.2 Economical Solar Energy**

The amount of solar energy received on the surface of the earth in less than an hour and a half is more than the yearly energy consumption of the world.[21,22] In an attempt to harness even a small fraction of this energy, some of the most attractive applications of semiconductors and their heterojunctions within recent decades have focused on harvesting solar energy to generate electricity, or to drive electrochemical processes (photoelectrochemistry). Of the various versatile areas of semiconductor applications in photo-science, the direct generation of electrical energy using solar cells remains the most researched and developed field. While all technology developed in this field shares a common goal, there exist a broad range of techniques, each based on different mechanisms for harvesting electricity from sunlight.[23,24]

It is estimated that the world energy consumption rate will more than double between 2001 and 2050 (from 13.5 TW to 27.6 TW) – and by the year 2100, will more than triple, reaching 43.0 TW.[21] To support this growth without increasing the atmospheric $CO_2$ concentration, the world will need to derive nearly 15 TW from carbon-neutral energy sources by 2050, and nearly 30 TW by 2100. This implies a need to produce more carbon-neutral power by 2050 than was produced from all energy sources combined in 2001. Of the most common carbon-neutral techniques (such as those involving hydropower, ocean-related features, wind, geothermal, solar electricity, solar fuels, and solar thermal energy), the solar energy harvesting approaches have the highest potential for fulfilling the carbon-neutral 15TW and 30 TW goals (2050 and 2100, respectively). Unfortunately, the actual usage of solar approaches lay several orders of magnitude below most of the other carbon-neutral energy sources. This remains a significant challenge and opportunity that the global community must seek to overcome and capitalize upon.

An important consideration in the development of practical solar technology is cost. In 2001, the average price of solar power was approximately $10/Watt (W). Over the past decade, prices have dropped close to $2/W, leading to a 100x increase in the number of silicon-based solar panels installed in the US, with an installation capacity of 5 GW in 2013. This reflects a small fraction of the global solar capacity, which overtook 150 GW in early 2014. Based on a detailed analysis of all major photovoltaic markets, the IEA Medium-Term Renewable

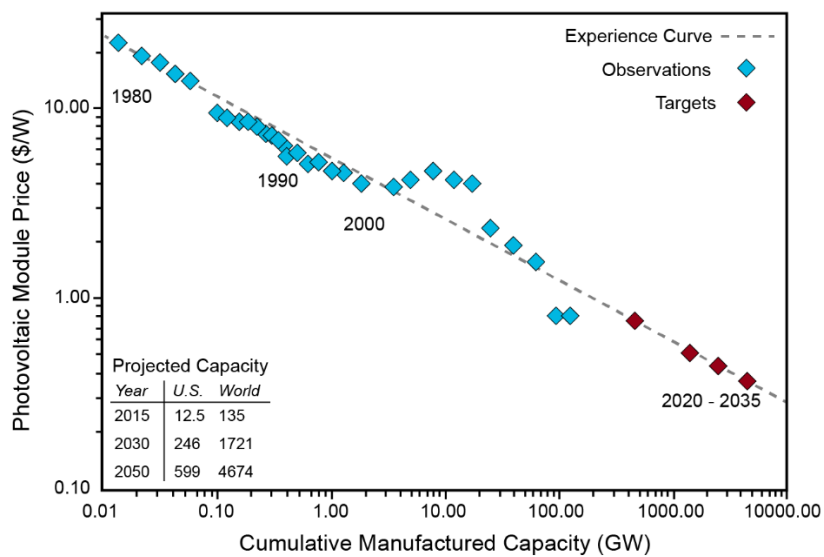

Figure 2: Global reduction in photovoltaic module price with projections up to year 2035, (adapted from reference 25 and 26). The inset shows projected an optimistic photovoltaic capacity (in GW) by the United States and the globe in 2030 and 2050.

Energy Market Report (IEA 2014c) conservatively estimates that cumulative installed photovoltaic capacity will likely exceed 400 GW worldwide by 2020.[25] China is predicted to lead the world, with over 110 GW, while Japan and Germany are each predicted to reach roughly 50 GW, with the United States following at over 40 GW. With respect to annual markets, by 2020, China will lead with about 14 GW/y, followed by the United States, at 5 GW/y.[25] Figure 2 shows an optimistic roadmap of the expected evolution of photovoltaic capacities of the United States cumulatively and the world over the coming decades, where the inset shows a projected capacity (in GW) of the United States and the globe in 2030 and 2050.[25,26] According to this roadmap, the global photovoltaic capacity at year 2050 will reach approximately 4.7 TW – roughly a third of the projected requirement from carbon-neutral sources in order to keep the $CO_2$ concentration fixed at the current level. This production will synchronize with the module price of photovoltaics, which is expected to drop to half its value over the next twenty years. Given the extensive resources already applied to the development of silicon-based (and other semiconductor) photovoltaic development, 2D materials are expected to play a niche role in this growing need for photovoltaic based electricity generation. A nanometer (or less) thick TMD can absorb 5-10% of sunlight, and can produce photocurrent both in the Schottky barrier as well as in the excitonic modes.[27] Owing to their wide range of optical gaps, exciton binding energies, work function values, mobilities, carrier types and densities, carrier lifetimes, and electrical resistivity, 2D materials form a natural system for development of complex heterostructures and junctions for optimized generation of solar power. The power conversion efficiency of 2D materials based solar cells have grown rapidly, from ~1% in 2013 to ~5% in 2014.[28] Several different 2D materials can be used for practical, high-efficiency applications.[29] A hybrid 2D-perovskite PV cell was reported in 2017, with best performance obtained at ~13%. While at this point (2018), power conversion efficiency of 2D materials are not expected to exceed those of conventional photovoltaic systems, their ultrathin form factor makes them highly suitable for incorporation into flexible and stretchable electronic systems. Hence, either 2D-only or 2D-enabled architectures are expected to provide new routes for low-cost energy-harvesting, especially in the domain of flexible and stretchable markets, where low cost of production will play an expanded role over time. **Low cost of manufacturing, improved mobility, scalable manufacturing and long-term chemical stability will be key metrics that will drive 2D materials into the future photovoltaics markets**.[30]

### 1.3 Internet of Things

In 2015, the White House administration challenged the nanotechnology community to "*Create devices no bigger than a grain of rice that can sense, compute, and communicate without wires or maintenance for 10 years, enabling an "internet of things" revolution.*"[31] The Internet of Things (IoT) is defined as a massive network of physical objects that sense and store information and communicate and/or act upon this information using algorithms enabled by embedded programming, big data analytics, and high-speed communication. This IoT challenge remains one of the biggest drivers for next generation connected reality. With more than a decade passed since the inception of IoT, several steps have been taken to create a foundation and sample of the bigger picture of IoT. For example, the initial stages of this movement can be traced to integration of radio frequency identification (RFID) tags in devices for tracking purposes, resulting in a network of connected devices. Since this development, IoT has found its place in a number of domains ranging from security applications, to transport, medicine and healthcare, as well as documentation applications in businesses. This has also resulted in reduced costs, enabling a broader scope for the improvement and implementation of IoT technology. Efforts in this field are now focused towards impacting people directly in their day-to-day lives by connecting daily objects, appliances, and gadgets. Additionally, efforts are underway to modernize technology to include positioning systems along with the tracking technology developed in the RFID phase of IoT efforts.

Broadly speaking, civilian IoT can be understood to have the most recognizable impact on wearable technologies as well as automotive, industrial, retail, farming, and healthcare industries, through the development of so-called "smart" technologies. Of these, the most sought-after applications from a consumer perspective appear to be smart home applications, including smart thermostats, lights, refrigerators and door locks. This is followed by wearables such as smart watches, activity trackers and smart glasses. Connectivity at the city and grid-scale draw the next level of interest, with applications related to smart parking, waste management, and metering, followed by industrial internet (remote asset control), connected cars, and connected healthcare technologies. At this point, smart retail, smart supply chain, and smart farming are at a level of curiosity.[32] Further market research shows that IoT will take 5 to 10 years to gain mainstream adoption.[33,34] According to market research predictions, the total number of IoT connections will grow from 6 billion in 2015 to 27 billion in 2025,[35] reflecting a compound annual growth rate of 16%.[36] This will be accompanied by data generation of 2 zettabytes, and an economic impact forecast of 2-5 trillion dollars.[37]

2D materials, due to their ultrathin form-factor, are quite well-poised to address several areas of the IoT hardware platform especially for ultra-thin and flexible/stretchable use-cases. In terms of active components, 2D materials have begun to show promise as switches, photodetectors, sensors,[38–40] and piezoelectric materials[41] – intrinsic components of the IoT hardware platform. In general, these atomically thin electronic materials will find niche applications in electronics, sensors and transmitters, and coatings among several other technologies. Hence, **in addition to the parameters that will enable them to serve high-performance computing (*e.g.* mobility, sheet resistance, chemical stability, scalability *etc.*), upon sufficient maturation, 2D materials will need to cater to the mechanical robustness required for flexible/stretchable/bendable technologies.**

### 1.4 Health

2D materials have attracted great attention in biosensing for healthcare applications. **2D materials offer high sensitivity due to large surface area, atomic thicknesses for optimized electrostatic modulation, tunable electronic/optical properties, flexibility, mechanical strength, and optical transparency.** The distinct chemical and physical properties of 2D materials make them ideal for detecting various biological targets, such as nucleic acids (ssDNA, dsDNA, and RNA),[42] proteins (different cancer biomarkers, antigens, antibodies, etc.),[43] and small molecules (glucose, $H_2O_2$, dopamine, lactate, ascorbic acid, etc.).[44–47] Various biosensor technologies have been developed based on 2D materials. For example, due to fluorescence quenching abilities of TMDs, a series of highly sensitive optical biosensors have been developed for detection of DNA and proteins.[48–50] Enhanced charge transfer rate as a result of surface defects and exposed active sites on the surface of 2D materials has attracted great attention in the development of electrochemical biosensors, especially non-enzymatic sensors. Importantly, the performance of 2D material-based electrochemical sensors can be further enhanced though combination with other nanomaterials, including noble metals, transition metals, carbon-based structures, and conductive polymers, resulting in synergistic effects for signal augmentation.[51,52]

Among various 2D materials, graphene, graphene family, i.e. graphene oxide (GO) and reduced graphene oxide (rGO), and their hybrids with other nanomaterials have attracted the greatest attention for biosensing applications. High conductivity, relatively simple functionalization with biorecognition elements, tunable optical properties, extraordinary mechanical strength, and biocompatibility are some of the unique properties that make graphene-related materials highly suitable for various biosensing techniques (electrochemical, field-effect based, optical)[53–56]. For electrophysiology, *in vivo* imaging, and manipulation of neuron cells (i.e. optogenetics) have been demonstrated.[57] Although most biosensing reports have focused on the graphene family and to some extent $MoS_2$ and $WS_2$, the potential of other 2D materials have not yet been explored. Among the unexplored groups are 2D materials doped or intercalated with reactive elements (e.g. Fe, Cu, Co, etc). These 2D materials could enable new avenues in biosensing, for example for real-time detection of trace levels of redox-active metabolites (such as reactive oxygen and nitrogen species), monitoring reaction kinetics and detection of extremely unstable complexes involved in various reactions (such as the Fenton reaction).

To achieve these 2D-based biotechnologies, simple, reproducible, and cost-effective synthesis methods must be established. Furthermore, control over layer number, size, defect density, and phase are required, as all of these characteristics influence sensing performance. Moreover, the lack of understanding and control of functionalization of 2D materials (for example with biolinkers used for immobilization of antibodies) limits their applications in biosensing. This is important, as functionalization plays an important role in stabilizing 2D materials in ionic solutions as well as tuning their electronic/optical properties. Additionally, biocompatibility, stability, and potential toxicity of various 2D materials need to be studied carefully. And finally, the mechanisms of charge transfer between 2D materials and various biomaterials (such as antibodies, bacterial cells, and nucleic acids) should be inspected systematically.

Addressing these technological aspects can open up new opportunities in future biosensing, such as spatiotemporal mapping of microbiome interactions at both single cell and community levels, analyzing target molecules beyond the current analytes, developing atomically thin, all-integrated sensors and circuit systems especially at the point-of-care (e.g. in flexible bioelectronics, implantable devices, and wearable technologies for continuous monitoring of physiological factors), and multiplexed detection of various biomolecules for developing new classes of diagnostics and therapeutics.

## 2. Synthesis of 2D Materials | *Techniques, Nucleation & Growth, and Substrate Impacts*

This section focuses on challenges and milestones related to synthesis of 2D materials which must be addressed to achieve electronic-grade material. It introduces key considerations related to growth techniques, nucleation and growth, and substrates, and the importance of these areas in the ultimate realization of electronic-grade material.

### 2.1 Powder, Gas, and Molecular Beam Source Techniques

> **Milestones:**
> - Realization of single-crystal, wafer-scale 2D films
> - Integration of TMD growth into silicon-based platforms, including amorphous oxide substrates and 3D structures.

Powder vaporization methods, also referred to as powder-based chemical vapor deposition (P-CVD), have been widely used to synthesize crystalline TMD domains and coalesced monolayer and few-layer films.[58,59] This straightforward method involves the placement of source powders and substrates in a quartz tube furnace and subsequent deposition of crystalline films on the nearby substrates (Figure 3a).[60] However, the source concentrations cannot be independently controlled and modulated with this technique, which ultimately limits the uniformity of the grown film over large areas, thus restricting its utility simply to "bench science". Gas source CVD is another technique for the synthesis of 2D materials, and holds particular promise for the realization of large-area films, due to the ability to control domain sizes and density through the use of precursor switching and pulsing schemes (Figure 3b).[61] This method utilizes volatile chemical precursors that are located outside of the deposition chamber and are controllably introduced as gases into the reactor. Gas source CVD, ubiquitous in semiconductor manufacturing, offers distinct advantages over competing 2D synthesis methods for process scale-up and high throughput – and will be a critical technique as the technology moves toward commercial applications. Multi-wafer CVD tools with robotic cassette-to-cassette sample loading and in-situ metrology, currently employed for Si and III-Vs, can be adapted for 2D layered chalcogenides with guidance from process simulation and reactor design considerations. Gas source CVD is also used in roll-to-roll processing for large-scale flexible electronics, and is proven as a robust technique to synthesize layered chalcogenide compounds including the tetradymite-type crystals[62–64] ($A_2B_3$ where A=Bi, Sb and B=Se, Te), transition metal dichalcogenides[65–68] ($MX_2$ where M=Mo, W, Nb, etc. and X=S and Se), group III[69] and group IV chalcogenides such as GaSe,[70] $In_2Se_3$[71–73] SnS,[74] $SnS_2$,[74] etc. As a result of the flexibility and control of precursor supply provided by gas source CVD, there has been renewed interest in its use for the synthesis of monolayer and few-layer chalcogenide films such as $MoS_2$[75–77] and $WSe_2$.[78–81]

A variety of precursors have been employed for the growth of 2D materials including metal carbonyls ($W(CO)_6$[75,80] and $Mo(CO)_6$[75–77]), halides ($MoCl_5$,[66] $WCl_6$,[65,78] $NbCl_5$[68]), metalorganics (($CH_3$)$_3$Bi,[62,82,83]), organo-chalcogen compounds (($CH_3$)$_2$Se,[80] ($C_2H_5$)$_2$Se,[65,78,82,84] ($C_4H_9$)$_2$Se,[66,68] ($C_2H_5$)$_2$S,[75] ($C_4H_9$)$_2$S,[77] ($C_2H_5$)$_2$Te,[83,85] ($C_3H_7$)$_2$Te[62]), and hydrides ($H_2S$,[76] $H_2Se$[80]). Except for the hydrides which are supplied as gases, the precursors are typically liquids or solids at room temperature with low to moderate vapor pressure. Precursor supply is controlled via the use of temperature and pressure-controlled bubbler manifolds which utilize a carrier gas such as $H_2$ or $N_2$ to transport controlled molar amounts of the volatile precursor into the deposition chamber. Precursor purity and chemistry can greatly impact the quality of grown materials and should be carefully considered when selecting appropriate sources for 2D material growth. For example, 90.0%-pure Se sources used for metal-organic CVD (MOCVD) growth of $WSe_2$ have been shown to result in carbon incorporation in synthesized films. Increasing the Se source purity to 99.99% is shown to eliminate this incorporation.[79] Additionally, the use of carbon-containing Se and S sources such as ($CH_3$)$_2$Se,[80] (t-$C_4H_9$)$_2$S, and ($C_2H_5$)$_2$S[84,86] can result in the formation of a defective carbon layer at the substrate-TMD interface. If carbon-free sources such as $H_2Se$[80] and $H_2S$[86] are used instead, this unwanted carbon layer is eliminated.[80] When developing growth schemes via CVD, deposition chamber-type should also be considered. Deposition chambers for gas source CVD can be classified into two main categories: hot-wall and cold-wall. In a hot-wall system, a resistive tube furnace is typically used to heat both the chamber walls and the substrate, while in a cold-wall system, only the substrate is heated and the walls are maintained at a lower temperature to minimize gas phase pre-reaction of sources and preventing upstream deposition on reactor surfaces, thus reducing contamination and allowing for the growth of multi-layers and heterostructures.[61]

Beyond CVD-based techniques, molecular beam epitaxy (MBE) also serves as a promising route for 2D material synthesis, particularly in the context of research and development, due to the potential for enhanced purity provided by a combination of high purity elemental sources and growth in an ultra-high vacuum (UHV) system. MBE utilizes ultra-pure sources which are heated in effusion cells, resulting in "beams" of atoms or molecules, which are then deposited on a heated substrate (Figure 3c).[87] The lower temperature growths in MBE also help in minimizing chalcogen vacancies and intermixing of layers during the growth of vertical heterostructures, which are common problems associated with high temperature growths in P-CVD and gas source CVD. MBE growth of transition metal dichalcogenides was investigated prolifically by Koma beginning in the 1980s.[88–90] Primarily focusing on the selenide family of TMDs, Koma and co-workers demonstrated a wide variety of TMDs grown on many van der Waals substrates[91,92] and pseudo-van der Waals substrates (passivated 3D substrates[93–95]). The interaction of the MBE-grown TMD with the underlying substrate was studied in detail by utilizing in-situ characterization methods including RHEED, auger spectroscopy, and scanning tunneling microscopy. Koma's group was the most active in the field of van der Waals epitaxy until 1994, when the Jaegermann group forayed into this field using metal organic MBE (MOMBE).[96] Interest in the epitaxial growth of TMDs plummeted around 2004, but interest rejuvenated in the 2010s, prompted by the potential application of TMDs as channel materials in advanced logic devices.[97–101] There have since been a number of TMDs demonstrated by MBE, including $WSe_2$[102], $MoSe_2$[101,103], $HfSe_2$[97], $ZrSe_2$[104], $SnSe_2$[99], $WTe_2$[105], $HfTe_2$[106], and $MoTe_2$.[107]

Due to the high vapor pressure of sulfur, the MBE growth of TMDs has primarily focused on the selenides and tellurides with a few rare exceptions.[108,109] This expanding family of TMD materials has accelerated the research field as the applications have expanded from logic devices, to spintronics, optoelectronics, magnetic devices, and superconductors.

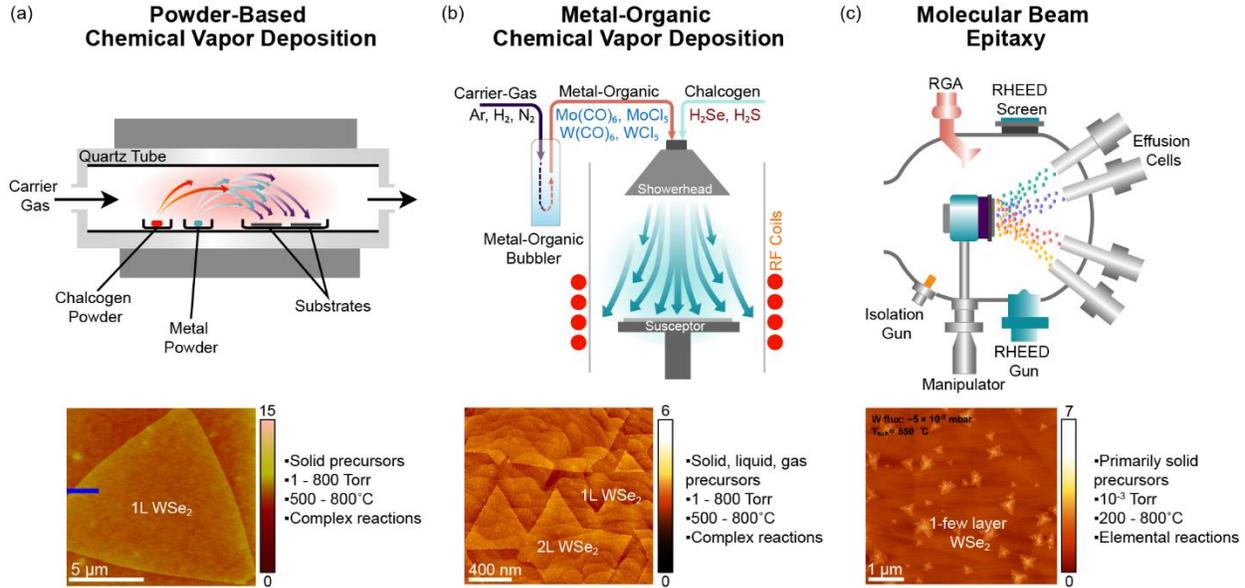

Figure 3: Overview of synthesis techniques and examples of mono and few layer WSe$_2$ produced by each. (a) Powder-CVD schematic and corresponding AFM image of WSe$_2$ layers on a SiO$_2$/Si substrate (adapted from reference 59) (b) Metal-organic CVD schematic and corresponding AFM image of mono and bilayers of WSe$_2$ on a sapphire substrate (adapted from reference 60). (c) MBE schematic and corresponding AFM image of few-layer WSe$_2$ on a highly-ordered pyrolytic graphite (HOPG) substrate (adapted from reference 86). AFM height scale bars

In recent times, it has been shown that the direct chalcogenization of Mo in vapor phase using MoO$_2$ as a source leads to samples with optoelectronic grade quality which is important for applications that harness their optical and exciton-based properties.[110–116] While gas source-CVD and MBE have been used to deposit a variety of chalcogenide thin films, extension to the controlled synthesis of monolayer and few-layer films and heterostructures presents new challenges. Specifically, reactors used in CVD techniques must be designed, and growth parameters chosen with the aim of achieving precise control over material uniformity and reproducibility. Substrates used for CVD growth must be thermally and chemically stable in the CVD growth environment and are also known to impact nucleation and film orientation. As the field moves toward electronic applications, process scale-up and throughput considerations are becoming increasingly more important. These issues provide guidance for future research and development activities. One of the most important and under-explored challenges is the integration of TMD growth on Si-based platforms. Allowable thermal budgets need to be critically assessed as many applications for TMDs are envisioned as back-end-of-the-line (BEOL) technologies with maximum temperatures of ~500 °C. To this end, a detailed understanding of 2D crystal quality as a function of growth temperature is required, along with methods to realize single-crystal material at temperatures of 500°C and below. Efforts in this area are relatively young and require continued investigation. In addition to low-temperature growth, one key challenge in the integration of 2D materials with Si platforms remains direct growth of chalcogenides on silicon-based substrates. Novel strategies for controlling reactivity of Si substrates in chalcogen environments are required if direct growth of TMDs on Si-based substrates is to become a reality. Along with this, single-crystal TMD growth on amorphous oxide substrates is also essential, requiring strategies for controlled nucleation of TMDs. Additionally, TMD growth on 3D structures remains an important area of study (FinFETs). Pending progress in the area of direct-growth, transfer-based approaches must be developed and assessed for feasibility and ease of integration with existing CMOS processes.

## 2.2 Modeling Gas Source-CVD Growth Chambers

**Milestones:**
- Computationally efficient, high-performance & high-fidelity models to inform real-time control of 2D material growth
- Computationally-informed reactor designs optimized for uniform growth conditions over large areas
- Growth chamber database to reduce 2D material production time

Although CVD and MOCVD have been widely used for the synthesis of several 2D materials, reproducibility and precise control during the growth process remain challenging, due to the complex physics involved in the many aspects of growth processes, including heat transfer, fluid flow, mass transfer, and chemical reactions. The field of 2D material growth faces challenges which are different and absent from previous extensive studies of thin-film growth, due to atomic-scale thickness of grown layers, morphology-dependent characteristics, and sensitivity to multi-length/time scale growth processes. Slight changes in growth conditions can drastically alter the final structure and characteristics, and key parameters of growth may be derived directly from the geometry of the macro-scale growth chamber and all the way down to the atomic-scale elementary processes that constitute growth in desired modalities, directions and structures. Physics-based, multiscale computational models of growth chambers are uniquely capable of providing insights into growth mechanisms, control over growth outcomes, and active control of growth conditions (pressure, flow rate, temperature, precursors) as guided by *in situ* monitoring.

Current growth chamber models adopt four major approaches: (1) the rate equations are parameterized experimentally,[117] (2) simplified analytical governing equations are used,[118] (3) an adaptive model is experimentally trained to determine the optimum growth parameters,[119–121] and (4) the coupled system of equations governing the growth processes at different spatial and time scales is numerically solved.[122–127] Models using the first approach require several trial experiments, where extrapolation to different experimental setups and beyond the specific experimental conditions is difficult. Although the second class of models provides the flexibility for adjusted experimental conditions, it is commonly oversimplified. The third class of methods requires a set of experiments to build the training set and its accuracy is determined by the number and diversity of the training experiments as well as the adaptive method itself. These models work well within the phase space of their training set but should be cautiously used beyond the conditions in which they were trained. The last class of these methods is intricate and requires significant computational resources. Transferring information between different scales and validation may become a challenge, and carefully controlled experiments will be needed to extract model parameters. However, these models provide a deeper insight into the growth process which cannot be gained from any of the other methods. They can also be adjusted to different growth conditions and be used to optimize the growth process.

A high-fidelity model of the MOCVD grown 2D materials should capture the formation and dissolution of chemical species as well as the heat and mass transport within the furnace. This involves accurately identifying gas-phase and surface reactions and growth, as well as their corresponding energy barriers and rate coefficients. This information can be obtained from atomistic simulations[128] or experiments such as mass spectroscopy.[129,130] The surface reactions involved also include chemisorption of gas-phase species and their reactions, desorption, and growth of the 2D material (Figure 4a).[126] Knowing the gas-phase reactions, some of the parameters for reactions can be estimated. For example, the maximum chemisorption rate of a species can be assessed using the collision rate of its gaseous phase with the substrate.[131] For conditions at which the concentration of reactive species is low, the change in pressure and heat of reaction can be neglected, and thus the fluid and heat transfer of gaseous materials can be decoupled from the mass transfer and kinetics. However, if the concentration of reactive species in the gas phase is high, e.g. in MOCVD, the complete set of coupled equations needs to be solved.[132–134] In addition to the sensitivity of 2D materials to subtle changes in thickness, morphology, and multi-length/time scale growth processes, challenges in growth of 2D materials include but are not limited to the versatility of synthesis methods, e.g. pressure control vs. flow rate control of the precursor.

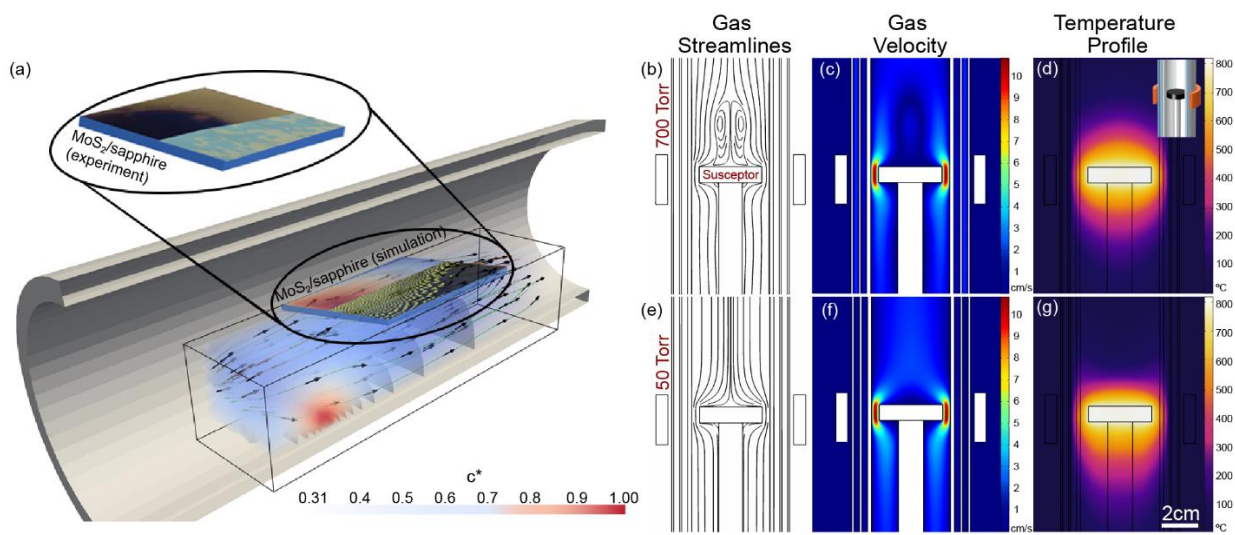

Figure 4: (a) schematic showing the concentration profile of $MoO_3$ in a tube furnace, and the resulting simulated and experimental distribution of $MoS_2$ deposition on the growth substrate (adapted from ref 125). (b-d) simulations of a cold-wall MOCVD reactor at 700 Torr, where (b) and (c) show the gas streamlines and velocity profile for 450 standard cubic centimeters per minute of $H_2$, and (d) shows the temperature profile for the susceptor and growth chamber. The inset depicts the susceptor, growth chamber, and surrounding RF heating coils. (e-f) gas streamlines, gas velocity, and temperature profile for the same reactor at 50 Torr.

Current CVD/MOCVD processes for TMD monolayer films typically involve high reactor pressure and low gas flow velocity to maintain a high partial pressure of chalcogen species above substrates. This serves to counteract the high volatility and high desorption rates of chalcogen sources during TMD growth. However, these conditions may lead to several problems that reduce the controllability of the CVD system. Figure 4b shows simulations of gas streamlines throughout a cold-wall MOCVD reactor, where buoyancy effects can be observed near the susceptor surface at high reactor pressures and low gas velocities. Buoyancy effects result in backflow, upstream deposition, and depletion of precursors, all of which can be detrimental to control over and ease of TMD growth. Additionally, high reactor pressure reduces the mean free path of precursor species and low gas velocity increases the gas residence time, both of which are likely to increase the possibility of gas-phase pre-reaction of precursors before they reach the substrate. Although these phenomena may not affect the current CVD/MOCVD processes using precursors such as $Mo(CO)_6$ and $W(CO)_6$ – as they will not lose all the CO ligands before reaching the substrate surface – other precursors such as TMGa and $H_2Se$ for GaSe films can have significant gas phase pre-reaction due to the Lewis acid and base properties of the two precursors[135,136]. In contrast, high gas flow rate and low reactor pressure can reduce the gas recirculation (Figure 4e) and precursor pre-reaction, however, this will also lead to change in temperature profile (Figure 4d, g) and reduction in precursor partial pressure.

Looking forward, several issues need to be addressed before a growth chamber model can become industrially relevant. The model should consider the relationship between the morphology and growth conditions, as the morphology directly affects the characteristics of grown layered materials. The **developed models should also have high-fidelity and high-performance**; they should be computationally efficient, especially when pursuing real-time active control of growth conditions. The **developed models should also facilitate the design of chambers which maintain uniform growth conditions over a large area, which is key to the reproducible synthesis of high quality 2D materials at an industrial-scale**. There are two main strategies for obtaining uniform growth conditions on a large scale: (i) an open-loop design approach, where the growth chamber is designed using the developed model for operation at a specific set of pre-defined conditions (and thus demand highly precise modelling); and (ii) the closed-loop design, in which the developed model not only can be used to design the growth chamber in the first place, but can also be employed to actively control the growth process during the experiment. In both cases, a precise growth model is vital. However, the first design approach is more sensitive to accuracy of the model, while computational efficiency is key in the second approach. The ability to actively control growth processes will be a new paradigm for designing 2D materials with novel structures and properties. The other research drive is to develop the **growth chamber database and charts to reduce the design to production time of 2D materials**.

### 2.3 Nucleation and Growth of 2D Materials

**Milestones:**
- Experimental control over nucleation and layer number of 2D films
- Computational representation of complex synthesis environments & corresponding impact on 2D material nucleation and growth

In addition to the orientation of domains forming a 2D film, it is **essential to actively control the number of layers deposited using thin film deposition methods**. Due to the chemical anisotropy inherent in the crystal structure layered materials, the growth rate of the edge is higher than that of the c-plane van der Waals surface. This gives rise to the planar morphology that is typical of layered chalcogenides. Hence, in an ideal scenario, it should be possible to nucleate and laterally grow domains of monolayer TMDs without significant growth of secondary domains on top. In practice, however, this is difficult to achieve due to the presence of defects on the van der Waals surface of TMD monolayers which act as nucleation sites for multiple additional layers.[137] Recent progress in this area has been made via manipulation of metal:chalcogen ratios and delineation of strategic growth steps in MOCVD.[79,81,138,139] However, for further improvement, *in-situ* characterization techniques are needed to provide real-time insight into nucleation and growth. The higher pressures used in CVD chambers render electron diffraction techniques ineffective; consequently, optical methods such as spectroscopic ellipsometry, laser reflectance, infrared absorption, and Raman spectroscopy are potential candidates for *in-situ* characterization. Within the growth chamber, adsorbants/surfactants provide a possible route to block the growth of secondary layers. The wide range of precursor chemistries available for gas source CVD may also prove advantageous for controlling the layer-by-layer growth process. It is challenging to achieve a digital layer thickness using MBE, mostly due to increased nucleation caused by metal-metal adatom interactions on the TMD surface.[87] High nucleation rates also result in small grains and increased number of grain boundaries, which degrade carrier transport.[87]

Substrate defects, like step-edges formed during annealing, surface vacancies or pits, also enhance nucleation and the growth rate, which promotes undesirable vertical growth. Epilayer-substrate interactions have also been observed through substrate defects,[138,140] and some preliminary evidence suggests chalcogen scavenging from a TMD or topological insulator substrate.[105]

To aid in efforts to control nucleation and layer growth, theoretical studies can be performed. Computationally representing the complexity of experimental syntheses environments remains a great challenge in materials research. Advances in molecular dynamics (MD) that treat the nuclei classically while considering the quantum nature of electrons will help these efforts by combining the strengths of methods such as Ehrenfest dynamics (offering the correct short-time coherent evolution at the cost of long-time deviations along the averaged-state trajectory) and surface hopping dynamics (whose stochastic branching over trajectories recovers the correct long-time behavior). Additionally, to address the relatively small time-scales sampled in MD, several advanced sampling tools have been developed,[141] which use modified potential energy surfaces that systematically reduce key reaction barriers[142] or use parallel replica concepts to essentially parallelize time.[143,144] Alternatively, combinations of Monte Carlo (MC) and MD methods can be used to improve the efficiency of MD sampling (ex. force biased MC method[145]). In addition to the atomistic models of the growth, other methods based on sharp-interface[146] approach and diffuse interface phase-field methods[147–149] have also been utilized.

Additional improvements in theoretical modeling can be sought through 1) scalable methods that treat a key subset of nuclear degrees of freedom at a fully quantum level and can describe kinetic processes involving light elements, 2) tight-binding methods which capture charge transfer, chemical reactions, and hybridization effects in a highly efficient empirical framework through the use of electron dynamics and coupling to external fields, and 3) real-time methods of electron dynamics made to model environments with unpredictable populations of transient molecular species in low symmetry environments. In addition to these real-time methods, corresponding methods for nuclear degrees of freedom can enable the study of low-dimensional materials in extreme environments. Many of these approaches scale particularly well in 2D materials, which provide a test bed for promulgating algorithmic advances to more traditional 3D materials. Thus, atomistic simulations of 2D material nucleation and growth will require hybrid methods that combine the strengths of first-principles techniques with the computational efficiency of empirical reactive force fields[150,151] and experimental validation.

Further requirements for studies of 2D material nucleation and growth lie in accurate prediction of energy barriers as well as a robust handling of large systems (ex. grain boundary structures) and time-scales long enough to capture diffusion and reaction barriers. One promising approach to address such systems lies in computational methods that can be trained based on *ab initio* calculations but have a significantly reduced computational cost (RCC). These methods include pseudo *ab initio* tools like tight-binding DFT,[152] reactive empirical force field methods,[153–156] and *ab initio* derived neural network tools[157] – typically methods that can handle greater than 1000 atoms and time-scales exceeding 1 nanosecond with relatively modest computational expense.[158] A critical connection with data science and machine learning will allow researchers to define the deviation of the RCC methods from the *ab initio*-trained configurational space and will allow the RCC methods to systematically receive additional *ab initio* training through sampling of new events or configurational space. These methods can be improved through the use of modern experimental imaging and spectroscopic techniques that enable real-time data acquisition from materials that are undergoing profound structural transformations (growth, defect creation, phase transformation, etc.). However, due to signal complexity and bandwidth constraints, only a small subset of available data is typically recorded and processed.[159–161] The complete collection and real-time analysis of material transformations would enable transformative advances in the degree of control over materials growth and transformation. This requires drastic improvements in acquisition, storage, and extraction of meaning from multidimensional data sets, facilitated by theoretical simulations that offer feedback to experimental parameters in near real-time.

## 2.4 Substrate Impact on 2D Materials

**Milestones:**
- Elucidate the nature of bonding between grown 2D layers and substrates
- Develop substrate engineering strategies for optimal 2D-layer transport
- Develop predictive tools for quantitative evaluation of electronic and optoelectronic properties of combined 2D-layer and substrate systems

The choice of substrate for growth of 2D materials is critical, given its impact on fundamental mechanisms such as nucleation, surface diffusion, film orientation, and lateral vs. vertical growth. A primary challenge in the field is the synthesis of uniform, large-area, single crystal mono- and few-layer films over substrates in which epitaxy or templating is required to orient domains. Preferential orientation of TMD domains have been demonstrated on c-plane sapphire for $MoS_2$,[162] $WS_2$,[163] and $WSe_2$,[164] (Figure 5a) but large area, electronic-grade films require defect-free merging of oriented domains – an active area of investigation.[61,81,163] Additionally, controlling the surface energy of the substrate via thermal treatments can enable epitaxial growth of TMDs like $MoS_2$[2,162] and $WSe_2$.[165] However, growth on non-single-crystal substrates such as oxidized silicon, glass, flexible metal foils and polymeric materials is also desirable for applications in the realm of flexible electronics. This requires moderate-to-low growth temperatures for gas phase CVD apart from improved control over nucleation and film orientation through surface seeding and patterning approaches.

Molecular beam epitaxy has a long history of growing layered materials, where studies have shown that the best substrates for high-quality 2D layer growth are other van der Waals materials (Figure 5b,c) such as TMDs[87,105,166,167], graphene or graphite[97,168], topological insulators,[169,170] and insulating materials like hexagonal boron nitride or mica.[104,171] The inert surfaces of van der Waals materials enhance adatom diffusion, allowing for the formation of larger grains with a more compact grain morphology. Experimental results of epitaxially grown TMDs on these inert, hexagonal substrates show that TMD layers grow unstrained, with intrinsic lattice constants, and are free of misfit dislocations.[97,103,105] Van der Waals epitaxy enables the growth of vertical heterostructures with materials chosen for their electronic properties instead of crystal lattice structure. Additionally, the defect density (e.g. chalcogen vacancies and extrinsic impurities) in TMDs grown by MBE on van der Waals substrates is $< 1\times10^{11}/cm^2$ due to the use of high-purity elemental sources with a controlled flux in a UHV environment.[172] Another advantage of *in situ*, van der Waals heterostructures is the ability to achieve rotational alignment between different layers, which has been predicted to result in a substantial increase in the on-current of tunneling devices.[173,174] While van der Waals interactions between the materials are too weak to induce strain or misfit dislocations, they are strong enough to influence the rotational alignment of the epi-layer with the substrate and this is observed with nearly all van der Waals heterostructure growth. To date, the quality of MBE-grown 2D layers on non van der Waals substrates is inferior to those grown on van der Waals substrates (Figure 3b) primarily because of stronger chemical interaction, which limits adatom mobility and increases nucleation rates. The growth of $MoSe_2$, for example, has been demonstrated on F-terminated $CaF_2$ by several groups,[168,175] with worse epi-layer quality compared to growth on graphene. Similar results are found for growth on sulfur and selenium-terminated GaAs(111)[94,169] and sapphire substrates.[170]

Substrates also greatly influence the photonic, electronic, and optoelectronic properties of 2D materials via strain, charge transfer and dielectric screening. For instance, when $MoS_2$ is laterally strained by 1.8%, it exhibits a direct to indirect bandgap transition, evident by a linear red shift of PL peak position and monochromatic decrease of photoluminescence (PL) intensity.[3] Most significantly, substrates can induce significant charge exchange across the 2D/substrate interface that substantially modulates PL emission (Figure 5d),[176,177] carrier lifetime,[2] field effect mobility,[178] and substitutional doping efficiency of 2D materials.[179,180] Furthermore, substrates with high dielectric constants similar to $HfO_2$ can screen the columbic potential interacting with charged impurities, resulting in field effect mobility values of greater than 100 $cm^2$/Vs for $MoS_2$.[181] Beyond these basic properties, substrates also serve as key components in the realization of interesting physical phenomena, such as room-temperature plasmonic resonance in $MoS_2$ on $SiO_2$/Si nanodisk arrays.[182]

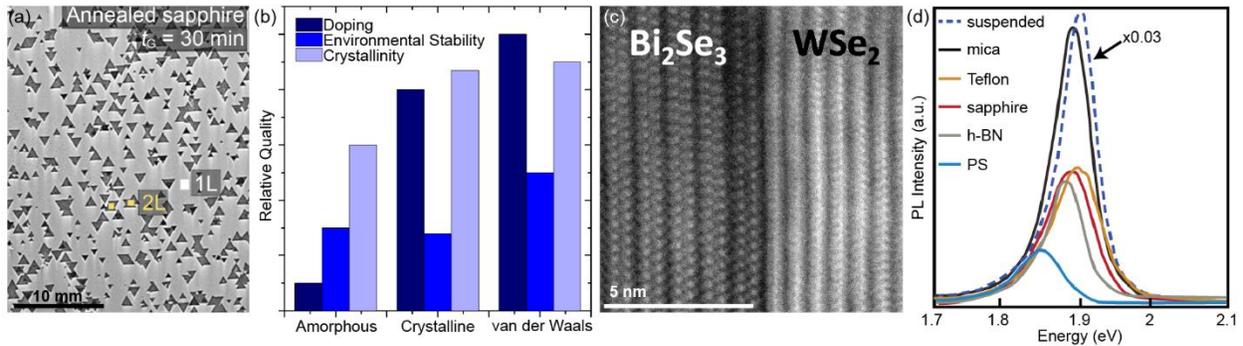

Figure 5: (a) Oriented $WSe_2$ domains on grown via MOCVD on an annealed, sapphire substrate (reproduced from reference 60) (b) relative comparison of ease-of-doping, environmental stability, and crystallinity of 2D materials grown on different types of substrates (c) cross-sectional transmission electron microscope image of $WSe_2$ grown on $Bi_2Se_3$ (reproduced from reference 86). (e) PL intensity measured for $MoS_2$ layers on different substrates (adapted from reference 176).

Substrates must also be carefully considered when computationally predicting 2D materials properties. For example, calculated band gaps of 2D materials will vary based on the presence or absence of substrates. Quantitatively accurate computation of band gaps from first-principles has been a major challenge for modern electronic-structure methods, and is particularly challenging in systems governed by fully non-local interactions, since such interactions involve the understanding of many-body effects. While great advances have been made in band gap computation through the development of non-local, hybrid functionals used in density functional theory calculations, such hybrid functionals do not capture non-local effects and band-gap renormalization that can be attributed to substrate screening. However, the use of greens function screened coulomb interaction approximation (GW) can provide a rigorous framework for accurate band gap calculations. The GW method has been applied to a wide variety of 2D materials, but presently available GW calculations performed on isolated systems significantly overestimate the band gaps compared to experiments when the material is supported on either metallic or insulating substrates. While the treatment of isolated adsorbates within GW is often feasible, a full treatment of the substrate is usually limited to either small molecules and correspondingly small surface unit cells[183,184] or thin supporting substrates.[185] An alternative is to incorporate screening from the substrate via classical image charge models,[183,186,187] which must be properly parametrized. Other approaches include constrained DFT and mapping to Anderson impurity models[188] as well as relying on hybrid functionals.[189] **The theoretical roadmap for 2D materials should therefore include the development of predictive tools for the quantitative evaluation of electronic band gaps that are both material (composition, structure, and size) *and* substrate dependent**. It is imperative to include the effects of a realistic dielectric environment, especially for 2D-based electronic devices such as field effect transistors. Such an approach could be based on the combination of methods where GW quasiparticle energies of isolated materials are corrected by accounting for classical screening of quasiparticle excitations.

Understanding the substrate impact and engineering the 2D/substrate interface hold unique promise to realizing electronic grade synthetic 2D films. **Deconvoluting the interaction between substrate and 2D material, and development of detailed understanding the interface properties are of vital importance to realize high quality synthetic 2D materials**. Specifically, the bonding type of synthetic 2D materials and respective substrates is still unclear. Many studies of 2D material growth indicate that the interaction between 2D films and substrate is beyond van der Waals bonding.[61,190–192] A combination of experiments and theoretical simulations to understand synthetic film-substrate interaction is necessary in making 2D materials technologically relevant.

## 3. Materials Engineering | *Defects, Doping & Alloying, and Heterostructures*

### 3.1 Defects

**Milestones:**
- Integration of multimodal, experimental and computational characterization to identify and understand defects in 2D materials and their impact on mesoscale transport
- Creation of a database to standardize defect characterization and engineering
- Experimental, atomic-scale control of defect structure, chemistry, and placement over large areas

Materials are inherently heterogeneous in composition and structure, through both localized defects such as vacancies or dopants and mesoscopic boundaries such as surfaces or interfaces, due to thermodynamic laws that favor fluctuation and disorder, especially in low-dimensional systems.[193,194] Irrespective of the synthesis method used, 2D materials contain several types of structural defects. The complexity and variety of defects present in 2D materials can also increase when considering the possible combinations of 2D alloying, interfacing and stacking of different material layers. Defects in these types of systems can break symmetries, scatter excitations, modify energy landscapes, and create quantum confinement. They often dictate material properties by controlling the creation, transport, and interconversion of excitations involved in electronic, optoelectronic, magnetic, thermal, or superconducting response,[195–199] and therefore, must be controlled to prevent degradation of material properties. Defects can be used extensively to control and tailor the properties of 2D materials for a variety of applications and also provides a platform to elucidate fundamental physical phenomena in 2D materials. As a result, the field of defect engineering is of great interest to the 2D materials community. Significant efforts have been devoted to identification, classification, and characterization of defects with various dimensionalities and structures in 2D materials.[151,200–207] Much of these efforts seek to understand how defects are introduced and formed in 2D systems, to develop post-synthesis treatments to generate artificial defects in a controlled manner, and to study how defects affect the properties and applications of the host materials. Overall, this field combines efforts to establish processing-structure-property-application relationships for defects in 2D materials.

The implications of defects in 2D materials can be both positive and negative. On one hand, defects can pose limitations for utilizing certain intrinsic properties of 2D materials such as their electronic mobility.[61]

On the other hand, defects provide opportunities to engineer 2D materials beyond the intrinsic limits.[208,209] As a result, it is important to controllably eliminate *and* generate defects. These two complementary goals will eventually allow scientists to combine efforts into a broader field of 'materials by design', in which materials with specific properties may be synthesized on demand. However, the field of defect characterization still faces several challenges, including: 1) integrated multimodal characterization to collectively examine, quantify, and identify defective samples, and 2) correlation of atomic-scale imaging and modeling, micro-scale spectroscopy, and mesoscale carrier transport. **The development of these complementary techniques in the context of 2D materials will enable the transfer of knowledge to wafer-scale systems.**

Research in defect engineering requires understanding and utilizing defects to realize 'materials by design'. To progress the field toward this end, structure-property relationships must be established with a two-fold goal in mind, where the primary goal is to discover new types of defective structures. Such new structures will continue to emerge with advances in synthesis and processing techniques that allow manipulating structures at atomistic level. For instance, one-dimensional tungsten chains have recently been observed in synthetic $Mo_xW_{1-x}S_2$ monolayers and[210] screw dislocations have been reported in CVD grown sulfide and selenide monolayers[211,212], nano-pores terminated by different edge reconstructions in $Mo_{1-x}W_xSe_2$ monolayers was triggered during *in-situ* heating in a scanning transmission electron microscopy (STEM), especially leading to a unique 2D-1D structure with the $Mo_{1-x}W_xSe_2$ monolayers terminated by metallic Mo(W)Se nanowire edge structure with potential novel properties[213]. As the family of 2D materials expands, new members such as $MoTe_2$, $ReSe_2$, ad $NbSe_2$ call for discoveries of unprecedented types of synthetic defects which may be related to a range of properties such as polymorphism and competing phases ($MoTe_2$), anisotropic in-plane bonding ($ReSe_2$), and rapid oxidation.[205,214–219] The second goal when investigating structure-property relationships is to understand new properties of defects. At present, most published results in the field address the impacts of defects on properties of semiconducting TMDs. Little is known about how defects affect the metallic, superconducting, and topological properties of other TMDs. It is imperative to establish the process-structure-property-application relationship via the development of innovative technical routes to control the types, locations, and concentrations of defects. Defect generation, migration, conversion, and elimination are topics of interest. As defective TMDs are more susceptible to chemical functionalization[206,220,221], defect-mediated surface modification is also a direction of tremendous potential. The additional focus of this phase is the exploration of novel applications enhanced or enabled by defects. To this end, sensing capabilities of 2D TMDs are expected to be expanded. TMDs functionalized with organic molecules or quantum dots may find novel biological and biomedical applications. To effectively progress toward defect engineering for materials by design, a **database regarding defect engineering must be established by researchers, based on which standardization of material qualities will be made possible**. These standards will provide a rational guide to industrial partners for commercialization of 2D TMDs. Defect engineering of 2D TMDs will eventually enter an era of "beyond 2D TMDs". Therefore, the properties of TMDs will be tailored on demand, and new functionalities will be enabled by artificial defects. Combining different types of defective TMDs (hybridize Mn-doped $MoS_2$ + Nb-doped $WS_2$, Se-deficient $NbSe_2$ stacked on $Mo_xW_{1-x}Te_2$), for example, will render the list of material possibilities nearly endless.

### 3.2 Doping & Alloying

> **Milestones:**
> - Establish rules for selection of dopants based on required properties
> - Controlled substitutional doping to $< 10^{10}$ dopants/cm$^2$ for tuning n- and p-type semiconducting layers

Dopants and impurities play significant roles in tailoring the optical and electrical properties of semiconductors. Within the field of 2D materials, several efforts have been made to dope and alloy 2D layers to tailor 2D material properties specifically. Most current efforts in "doping" 2D semiconductors are more appropriately categorized as alloying, due to the use of dopant concentrations well above 1%. Such concentrations constitute an equivalent doping of $10^{22}$ cm$^{-3}$ in silicon, which is many orders of magnitude beyond doping levels required for tuning Si electronic properties ($10^{14} - 10^{18}$ cm$^{-3}$). Nevertheless, doping and alloying of TMDs – primarily Mo- and W- based – is possible today due to relatively mature synthesis methods that enable opportunities for property tuning. As a result, a number of successful strategies for doping 2D materials have been developed, including direct charge injection via electrostatic gating,[222–225] charge donation from physically or chemically adsorbed molecules or ions,[226–231] and covalent bonding[232,233] via edge functionalization or substituted atoms.[4–8,234–238] These doping methods impact the optical, electrical, and optoelectronic properties of 2D TMDs. For examples, physisorbed or chemisorbed molecules on 2D layers act as molecular gates by injecting charge, which can modulate PL intensity through different charge carrier polarity.[227,228] Molecular gates can also modulate the carrier type and density in 2D semiconductors more effectively than conventional electric field gating, which not only enriches p-type and n-type building blocks for electronic and optoelectronic devices,[226] but also introduces metallic[224,229] and superconducting behavior.[225,230]

Many types of doping can be employed to engineer the properties of 2D materials. Substitutional doping in the 2D lattices can enable the formation of robust, stable alloys, where substituting atoms can be electron or hole donors to tune the free carrier type in semiconductors, e.g., Nb (hole donor)[7,239] and Re (electron donor)[180,240] in MoS$_2$, which can result in degenerate doping.[7] In addition, isoelectronic substitutional doping/alloying is another effective way to tune the electrical and optical properties of 2D TMDs, where dopants can more easily enable alloy formation due to good lattice matching with the host material. One example of property modulation via isoelectronic alloying can be seen from optical bandgap tuning of 2D TMDs via chalcogen substitution, e.g., MoS$_{2(1-x)}$Se$_{2x}$, or transition metal substitution, e.g., Mo$_x$W$_{1-x}$S$_2$, with 0<x<1. Tuning of electrical properties has also been demonstrated through alloying in a CVD grown monolayer Mo$_x$W$_{1-x}$Se$_2$ system, in which the substitution of Mo with W in monolayer MoSe$_2$ suppresses the characteristic n-type conduction and enhances p-type conduction with increasing W concentration, which becomes dominant with only ~18% of W substitution.[4]

Apart from band structure engineering, isoelectronic alloying of W in the Mo$_x$W$_{1-x}$Se$_2$ system can also significantly suppress the formation of Se vacancies and deep defect levels,[236,241] due to a stronger hybridization between the outermost p orbitals of Se and the d orbitals of W. As a result, Mo$_x$W$_{1-x}$Se$_2$ monolayers show ~10x more intense PL and an increase in the carrier lifetime by a factor of 3 compared with pristine MoSe$_2$.[236] More recently, isoelectronic alloying was also used for phase engineering in 2D TMDs. For example, 2H-MoTe$_2$ can transform to the 1T' phase at room temperature simply by W-substitution to form Mo$_{1-x}$W$_x$Te$_2$ alloys, where the critical "x" value needed to stabilize the 1T' phase is ~0.08.[5] Since the 1T'-phase MoTe$_2$ is predicted to possess unique topological properties, it is possible to produce high-performance optoelectronic devices with non-dissipative transport channels through the combination of semiconducting and topological elements in individual compounds by manipulating the local composition of $Mo_{1-x}W_xTe_2$ alloys.[5]

Despite many achievements in alloying of 2D semiconductors, there are still great challenges. Currently, very few atoms have been shown to successfully substitutionally exchange during CVD growth. For example, as predicted by theoretical calculations, dopant atoms such as Co, Fe, Ni, Mn, V, and Cr can introduce magnetic (ferromagnetic or antiferromagnetic) properties into 2D TMDs.[242–244] However, so far, the only reported doping of magnetic atoms in 2D TMDs is Mn doped in $MoS_2$ grown on graphene substrates – suggesting that the substrate may play a dominant role in the doping process.[179] Besides TMDs, it is highly expected that doping strategies can be developed for other 2D semiconductors. For example, GaSe, which belongs to the III-VI family of 2D layered semiconductors, shows many unique optoelectronic and nonlinear optical properties.[245–249] More attractively, theoretical calculations have predicted that monolayer or few-layer GaSe will show tunable ferromagnetism if the hole density can be significantly increased (to ~$3 \times 10^{13}$ cm$^{-3}$) – a great experimental challenge.[250] Other challenges include precise control of concentration and uniform distribution of the dopants, which will be necessary down to ~$10^{10}$ cm$^{-2}$ if we are to achieve CMOS-level doping precision.

The roadmap for doping in 2D semiconductors should be focused on solving these current challenges. Specifically, we **must first refine theoretical models to provide rules for the selection of dopants and design of alloys**, including considerations such as lattice matching, bandgap matching, formation energy, and bonding energy.[251] Second, **strategies must be developed for controllable substitutional doping at <$10^{10}$ dopants/cm$^2$**. This milestone is crucial for the development of versatile, 2D-electronics. Furthermore, methods are needed to substitute different types of atoms and inject more carriers in 2D semiconductors while decoupling the impacts of the substrate (see previous section). For example, recent advances in substitutional doping with magnetic atoms,[179] suggest that substitutional doping can be enabled through substrate engineering. Moreover, highly non-equilibrium synthetic processes, such as pulsed laser deposition (PLD), demonstrate a potential route for the growth, doping, and defect-manipulation of 2D semiconductors beyond that possible in CVD or MBE.[252,253] PLD, in combination with CVD, could be exploited further as new doping strategy for 2D semiconductors. It is also worth noting that the doping processes mentioned above involve introducing 'foreign' elements into the material lattice. However, the manipulation of the nuclear mass of the atoms composing the intrinsic material lattice should also be considered. One way to realize this doping is through isotopic modification, which not only influences phonon frequencies and phonon-related properties but also can influence electronic properties, such as electronic band structure and exciton binding energies due to electron-phonon interaction and coupling.[254,255] Therefore, isotopes represent a degree of freedom that could be exploited to tune the physical properties of the material at hand, while retaining nearly identical chemical behavior. Recently studies have demonstrated tunable phonon-related properties of graphene through modification of carbon isotopes.[256] In addition to graphene, theoretical calculations also predict unusual isotope effects on thermal conductivity of semiconducting $MoS_2$ monolayers.[257] Isotopic modification is also expected to affect the electrical, optical, and magnetic properties of 2D materials, which could represent a new pathway for developing new functionalities and devices based on 2D material systems.

## 3.3 Heterostructures

> **Milestones:**
> - Controlled synthesis of large area lateral and/or vertical van der Waals heterostructures with pristine interfaces and no intermixing of disparate layers
> - Predictive, computational efforts to inform the realization of novel heterostructures

The development of 2D layered heterostructures offers a nearly infinite number of potential combinations for tuning electronic, optical, chemical, and structural properties of the final materials. The earliest heterostructure works focused on the exfoliation of hBN as a substrate and an encapsulating layer to investigate the electronic properties of graphene, leading to record mobility reports and demonstrating that interfaces free of dangling bonds are key to accessing the intrinsic properties of these 2D layers.[258] This technique is now "standard" for exploration of intrinsic properties of 2D layers, with dozens of publications per year demonstrating that hBN encapsulation is essential for fundamental physics experiments. Beyond encapsulation, predictions of unique properties resulting from stacking different 2D layers together indicate that the stacked layers no longer act as individual components, but rather, interact to yield new properties not present in the constituent layers.[259,260] These findings have prompted robust efforts to fabricate various combinations of Mo- and W-based TMD heterostructures, including mechanical exfoliation and stacking of $MoS_2/MoSe_2$,[261] $MoS_2/WSe_2$,[262,263] and $MoSe_2/WSe_2$[264]. While mechanical exfoliation is key for rapid advancement of the fundamental physics of the heterostructures, it is also critical to understand and develop synthesis techniques capable of achieving the same structures over large area, if these materials are to be utilized "beyond the bench". To this end, there is a robust effort to synthesize such heterostructures (Figure 6), including but not limited to: graphene-based heterostructures,[265–270] hBN-based heterostructures,[270–273] and TMD-based heterostructures,[138,274–282] with many summarized in the large number of review articles published each year on the subject.

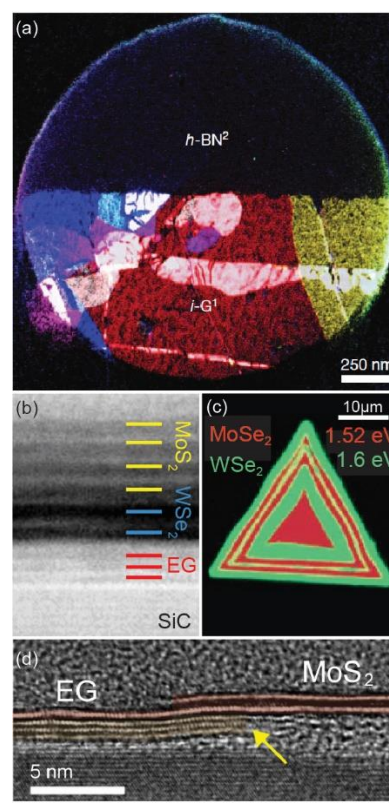

Figure 6: Examples of heterostructures composed of (a) lateral graphene/hBN, (b) vertical TMD/Graphene, (c) lateral TMDs, and (d) lateral TMD/graphene. Figure adapted from references 137, 269, 279, 281.

Unlike mechanical exfoliation, growth of heterostructures requires more cognizance from the researchers in understanding the growth method. In most cases, it is of paramount importance to not grow heterostructures without first understanding the chemical and thermal stability of the layers that act as the substrate for each subsequent layer. For instance, it is very difficult to grow graphene or hBN on a TMD substrate. This is because the temperatures required for high quality graphene[283] or hBN[284] growth far exceed the stability of the TMD layers,[285,286] which degrade rapidly via chalcogen (S,Se,Te) loss or metal oxidation at elevated temperatures. Furthermore, one must consider the exchange of chalcogen elements in TMD heterostructures. For instance, growth of $MoS_2$ on $WSe_2$ has been shown to be possible without S/Se exchange, while under similar conditions, the growth of $WSe_2$ on $MoS_2$ leads to the complete exchange of S by Se, forming $MoSe_2$.[138] Such exchange processes can be limited, but a thorough understanding of the kinetic and thermodynamic processes must be developed to provide a robust route to developing the desired heterostructures while precluding degradation or alloying.

Moving forward, it is essential to consider not only how to grow these layers but also understanding why to grow them. Many lateral heterojunctions, where two TMDs are grown from the edge of another TMD, argue that the formation of atomically thin and atomically sharp p-n junctions could provide a unique form of the junction. However, such structures do not perform as well as traditional "3D" p-n junctions, and therefore remain at the curiosity level for most of the community. Vertical heterostructures, on the other hand provide evidence that stacking of the layers can lead to unique electronic properties that benefit from pristine interfaces. One such example is the growth of TMD heterostructures on a graphene lattice that leads to resonant tunneling, with negative differential resistance curves that exhibit some of the most narrow resonant peaks reported.[138] Another example is that of TMD-graphene heterostructures where the graphene is probed as a contact material. It is shown that pristine interface of in this heterostructure leads to a reduction in the contact resistance.[280] Furthermore, 2D/3D hybrids also provide evidence that such structures could outperform silicon-based transistors. To realize heterostructures to their fullest potential, **techniques for controllable synthesis of heterostructures over large areas must be developed, in which pristine, interlayer interfaces are maintained**. Additionally, instead of growing heterostructures for the sake of making a new material, **it is important to understand the need to grow these layers by working closely with theory colleagues** to explore which structures (vertical vs. lateral or a combination of both, semiconducting vs. metallic) will lead to properties that are either a significant improvement on current technology, or that lead to completely new technologies.

## 4.  Outlook | *Electronic Devices from 2D Materials*

When considering the implementation of 2D materials-based devices for the electronics industry, it is important to think on the building blocks of our current electronic technologies. It took decades of research, development and most importantly billions of dollars of investment at the national and international level for Si technology to become the ultimate benchmark for all other electronic materials and devices. It is also true that, at present, the evolutionary path of Si technology, driven by Moore's Law of Scaling, seems to be narrowing and fast approaching an end simply due to the fundamental limitations of Si at the atomic scale.[16] High quality 2D materials are promising next-generation alternatives, but are at least a decade away from making inroads to the conventional semiconductor industry. In fact, it is unlikely that 2D technology will supplant Si; but instead may coexist *with* Si technology. This, too, will require significant research and resource investment in large area growth of 2D materials at temperatures compatible with silicon-based technology, including back-end of the line (BEOL) process flows. Even then, there is no consensus on what role the 2D materials and devices will play on a hybrid and heterogeneous Si platform. Interconnects, diffusion barriers, and access resistors for memory elements seem to be more acceptable for use of 2D materials than the replacement of logic units.[287,288]

2D materials, however, can be immediately incorporated into new technologies such as those composing the Internet of Things (IoT), where the thresholds for high-performance are far less stringent, and the demand for energy efficiency, multifunctionality, and low manufacturing costs are superior.[289] 2D devices in the form of chemical, biological, thermal, mechanical and optical sensors are not only attractive due to their high detection efficiency, but are also poised to be integrated into IoT technology,[290] flexible electronics and display electronics.[291]

Finally, beyond traditional electronic applications, high quality 2D materials appear to be at the cusp of a variety of breakthroughs in novel quantum phenomena, which will also have far reaching consequences. Devices based on many-body effects, valley physics, excitons and trions can offer a paradigm shift if harnessed properly.[292–294] Similarly, straintronic, piezotronic and mechatronic devices based on 2D materials can provide breakthrough and energy efficient solutions for computation, automation and communication if explored with persistence.[295,296] 2D materials can also enable platform for hardware artificial intelligence through neuromorphic devices which can mimic the human brain.[297,298] Heterostructures of 2D materials also offer endless opportunities for novel devices, where the primary challenge is to identify the right stacking or stiching.[299] These futuristic applications will, however, require long term vision and determined and sustained research efforts instead of single prototype demonstration.

## Acknowledgements

This roadmap has been a collaborative collation of information amongst several experts in order to establish a roadmap for 2D materials. Z.L and M.T acknowledge support from Simin Feng, Kazunori Fujisawa, and Xin Gan for their input with the roadmap. S.K and J.A.R also acknowledge support from Luigi Columbo for his insights on the 2D Roadmap. N.B, J.M.R, and J.A.R acknowledge the funding from the 2D Crystal Consortium National Science Foundation (NSF) Materials Innovation Platform under cooperative agreement DMR-1539916. N.B. and J.A.R acknowledge the Semiconductor Research Corporation Intel/Global Research Collaboration Fellowship (SRCEA Fellows and Scholars Program, task 2741.001). S.S and J.A.R acknowledge the NSF CAREER (Award: 1453924). K.Z and J.A.R acknowledge the Center for Atomically Thin Multifunctional Coatings (ATOMIC), sponsored by the NSF division of Industrial, Innovation & Partnership (IIP) under award #1540018. S.K acknowledges support from the NSF CAREER ECCS 1351424. A.E acknowledges support from the Pennsylvania State University Materials Research Institute. K.M acknowledges the support from LONI, NSF EPSCoR CIMM-OIA-1541079 (CFDA #47.083), LaSPACE- PO-0000028218, ATOMIC. X.L., D.G. and K. X. acknowledge the support by the U.S. Department of Energy, Office of Science, Basic Energy Sciences (BES), Materials Sciences and Engineering Division and the Center for Nanophase Materials Sciences, which is a DOE Office of Science User Facility. R.M.W and C.L.H acknowledge the support in part by the Center for Low Energy Systems Technology (LEAST), one of six centers supported by the STARnet phase of the Focus Center Research Program (FCRP), a Semiconductor Research Corporation program sponsored by MARCO and DARPA. It is also supported by the SWAN Center, an SRC center sponsored by the Nanoelectronics Research Initiative and NIST. R.M.W also acknowledges the support in part by NSF Award No. 1407765 under the US/Ireland UNITE collaboration.